\documentclass[prl,aps,twocolumn,superscriptaddress,showpacs,floatfix]{revtex4}
\usepackage{epsfig,graphicx,latexsym}
\usepackage{amssymb,amsmath}
\begin{document}

\title{\bf\noindent On the fluctuation  of thermal van der Waals  forces due to dipole fluctuations}

\author{David S. Dean}
\affiliation{Universit\'e de  Bordeaux and CNRS, Laboratoire Ondes et
Mati\`ere d'Aquitaine (LOMA), UMR 5798, F-33400 Talence, France}
\author{V. Adrian Parsegian}
\affiliation{Department of Physics, University of Massachusetts, Amherst, MA, USA}
\author{Rudolf Podgornik}
\affiliation{Department of Physics, University of Massachusetts, Amherst, MA, USA}\affiliation{Department of Theoretical Physics, J. Stefan
Institute, SI-1000 Ljubljana, Slovenia} 
\affiliation{Department of Physics, Faculty
of Mathematics and Physics, University of Ljubljana, SI-1000
Ljubljana, Slovenia}
\begin{abstract}
Fluctuations of the thermal or classical component of the van der Waals  force between two dielectric slabs, modelled as an ensemble of polarizable dipoles which interact via the usual electrostatic dipole-dipole interaction, are evaluated.  In the model the  instantaneous force is a deterministic function  of the dipole configurations in the slabs and its fluctuations are purely due to dipole fluctuations (no background thermal fluctuations of the electromagnetic field are considered).  The average of the force  and its variance  are computed. The fluctuations of the force exhibit normal thermodynamic scaling in that they are proportional to the area of the two plates, and even more importantly, do not depend on any microscopic cut-off in the theory. The average and the variance of the thermal van der Waals forces give a unique fingerprint of these fluctuation interactions.
\end{abstract}
\maketitle
\section{Introduction}
The Casimir force \cite{casgen} is the force arising between objects placed in a quantum and/or thermal field  due to the modification of the fluctuations of the field by the presence of the objects. As the force is due to a fluctuating field,  the force itself should  fluctuate. What is normally given as the Casimir (quantum or thermal due to the zero frequency  Matsubara mode) force is  its {\em average} value measured in thermodynamic equilibrium. Deriving a fingerprint of the average force and fluctuations around the average force could aid identification of the Casimir interaction component in an otherwise complicated experimental setup, showing long-range interactions of mixed origins \cite{Disorder}.

The first analysis of Casimir force fluctuations pertains to the quantum context \cite{bar1991, ebe1991}. In this case, despite the fact that the average value of the force is finite, the variance of the force exhibits an ultra-violet  divergence that is  eliminated using the experimental fact, that the force is always averaged over a time corresponding to the temporal sensibility of the experimental apparatus. From this analysis, while the average force on a single mirror is zero, its variance is non-zero, due to differences in the electric field on either side of the mirror. These fluctuations of the Casimir force can in turn be related to the force exerted on a non-uniformly accelerating mirror \cite{jac92}. In addition, the fluctuations of the radiation pressure exerted by a laser beam on a conducting surface can also be analyzed \cite{wu01}, in this case calculations using the electromagnetic stress tensor can be confirmed by a kinetic like approach based on photon number fluctuations. 
Casimir force fluctuations between perfectly conducting mirrors at finite temperature have also been examined \cite{rob1995} and it was shown that the high temperature force could be derived from the classical Rayleigh-Jeans 
distribution, while the force fluctuations required a full quantum treatment before taking the classical limit. In all the above cases average forces and their fluctuations are seen to arise via the boundary conditions imposed by the conductor on the electromagnetic field. In \cite{wu02,mes07} fluctuations of the Casimir-Polder interaction between a polarizable atom and a perfect conductor were  analyzed, presenting a conceptual departure from previous studies as force fluctuations can be related to a 
physical property of the atom, its polarizability, going beyond descriptions of objects in terms of ideal
boundary conditions.

In \cite{bar2002} the fluctuations of the thermal Casimir force due to a free mass-less scalar field theory with Dirichlet boundary conditions on parallel plates was considered. The leading term in the variance for two plates of area $A$  separated by a distance $L$ was shown to posses a cut-off dependent limit $\langle f^2\rangle_c \sim(k_BT)^2A/a^4$, where $T$ is the temperature of the system, $k_B$ is Boltzmann's constant and $a$ is a microscopic  (ultra-violet) cut off. The average value of the force in this system behaves as $\langle f\rangle \sim  k_BT A/L^3$. Thus in terms of the extensive variable $A$, the fluctuations of the force $\Delta f$ obey the usual thermodynamic scaling $\Delta f \sim \sqrt{A}$, compatible with the notion that local fluctuations of the force at distant regions of the plates are uncorrelated. 
Similar results have also been found for the fluctuation induced forces on inclusions, such as proteins, in membranes \cite{bib2010}.  

More recently \cite{rod2011} the fluctuation  of the Casimir force for scalar fields was examined in a parallel plate piston cylinder geometry, with the result that $\langle f^2\rangle\sim 2\langle f\rangle^2$, which is clearly at variance with the thermodynamic scaling found in \cite{bar1991, ebe1991,bar2002}. Whether thermodynamic scaling should hold is far from obvious as all the fluctuation induced interactions in the studies above are long range (corresponding to mass-less field theories). 

\begin{figure}[t!]\begin{center}
	\begin{minipage}[b]{0.312\textwidth}\begin{center}
		\includegraphics[width=\textwidth]{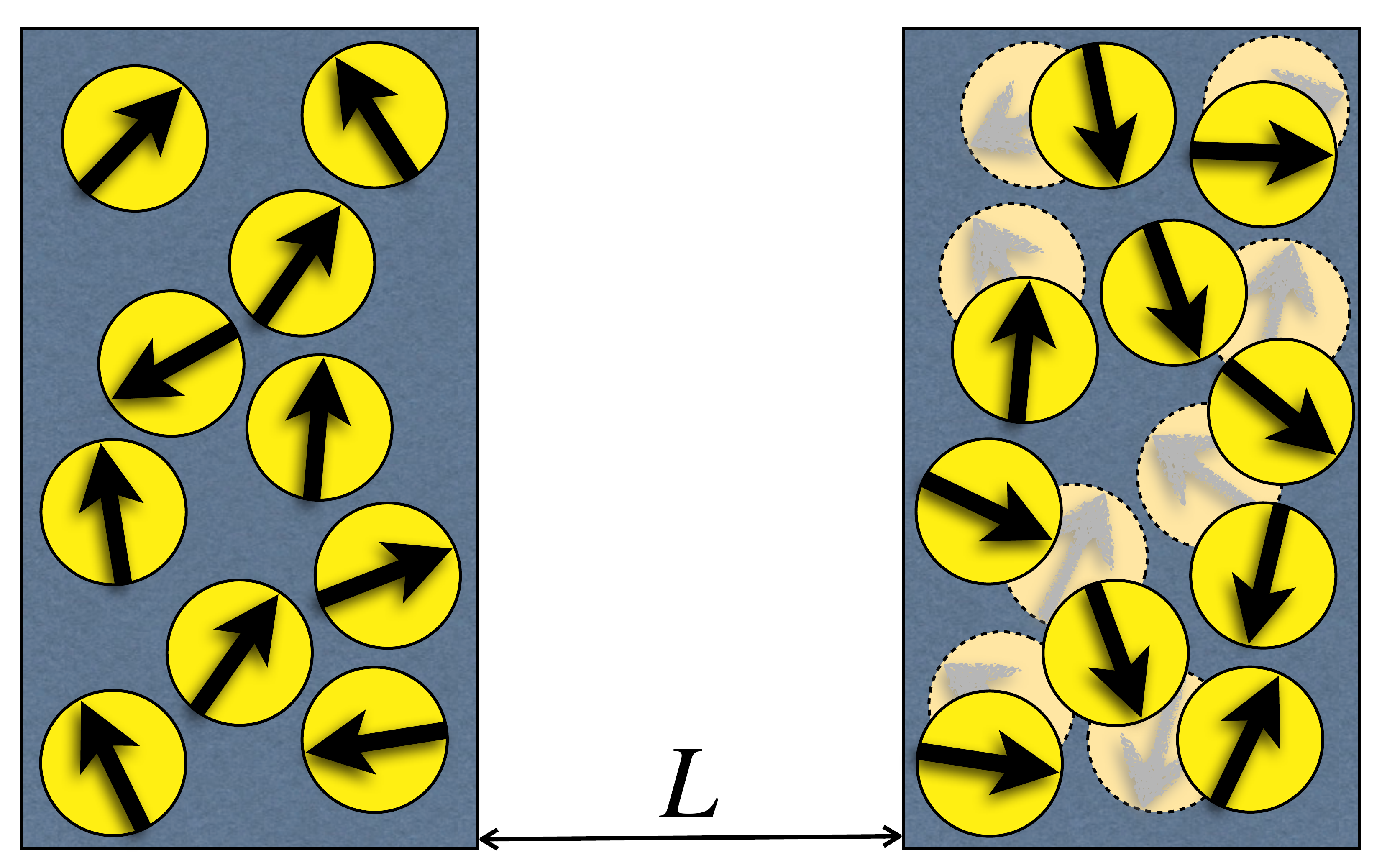}
	\end{center}\end{minipage} 
\caption{(Color online) Schematic drawing of a classical dipole model of two dielectric half-spaces at separation $L$. The slit region is a vacuum. Fluctuating thermal dipole configurations (shown bold and faint) create a fluctuating thermal Casimir interactions that for large $L$ decay to zero strictly with a probability one.}
\label{schema}
\end{center}\end{figure}

Connected with the question whether fluctuations in Casimir forces scale thermodynamically, is the intriguing existence of fluctuating forces on isolated bodies, depending on a microscopic cut-off, as such forces should induce movement/diffusion of small inclusions and perhaps also fluctuation-induced drag forces, which have been  predicted in a number of quantum \cite{drag} and thermal \cite{tdrag} situations. The theoretical results described above point to the existence of regimes, notably in microscopic systems, where fluctuations of measured Casimir forces of both quantum \cite{expsq} as well as thermal or critical Casimir provenience \cite{expscrit}, can become large and thus may be experimentally measurable.   

In \cite{dea2012} a model dielectric was introduced, based on a continuous polarizable dipole field, showing thermal fluctuation forces identical to the zero frequency Matsubara mode van der Waals forces of the standard Lifshitz theory with media of the same dielectric constants. Within this model it is possible to analyze straightforwardly how the van der Waals interaction arises from the correlations between dipoles in opposing slabs and how the van der Waals force evolves temporally when switched on from zero, that is for initially uncorrelated slabs, to its final equilibrium value. 

Here we use the very same model to study the fluctuations of the dipole-dipole induced thermal Casimir force in the problem. Because we analyze direct interaction between dipoles, it is clear that as the distance between the two slabs is taken to infinity, the force itself must go to zero. In our model we only take into account the electromagnetic field generated by the dipoles in the slabs and ignore fluctuations of the electromagnetic field due to thermal photons (which cannot be properly included in the classical limit considered here \cite{rob1995}). In this situation therefore, we do not expect a {\em bulk} cut-off dependent force fluctuation of the type mentioned above.  We find that the average force scales as $\langle f\rangle \sim Ak_BT/L^3$ while the variance behaves as $ \langle f^2\rangle_c \sim A(k_BT)^2/L^4$ and tends to zero for infinitely distant bodies. We note that the variance obtained behaves like the $L$-dependent part of the variance found in \cite{bar2002}.
 
 \section{Polarizable field model}
We consider a classical model of interacting dipoles introduced in \cite{dea2012}. Here we have two slabs of material separated by distance $L$ in the direction ${\bf e}_z$ and the Hamiltonian for the system is given by
\begin{equation}
H = {1\over 2}\int d{\bf x}d{\bf x}' \sum_{ij}p_i({\bf x})A_{ij}({\bf x},{\bf x}';L)p_j({\bf x}')
\end{equation}
where ${\bf p}({\bf x})$ is a local dipole field. The interaction is given by
\begin{equation}
A_{ij}({\bf x},{\bf x}':L) = {\delta_{ij}\delta({\bf x}-{\bf x}')\over  \chi({\bf x})} + D_{ij}({\bf x},{\bf x}';L),
\end{equation}
where $\chi({\bf x})$ is the local polarizability of the dipole field at the point ${\bf x}$ and $D_{ij}$ is the 
usual dipole dipole interaction. In slab $S_i$ we set $\chi=\chi_i$ (for $i=1$ and $2$) and we 
use a coordinate system such that the points in $S_1$ are in the half space $V^-:$  $z<0$ and the 
points in the slab $S_2$ are in the half space $V^+$: $z>0$. The dipole-dipole interaction is then given by
\begin{equation}
D^{(0)}_{ij}({\bf x}-{\bf x}') = \nabla_i\nabla'_jG_0({\bf x}-{\bf x}')
\end{equation}
where $ \nabla$ indicates the gradient with respect the the coordinate ${\bf x}$, $\nabla'$ the gradient with respect to ${\bf x}'$ and $G_0$ is the vacuum Green's function obeying.
\begin{equation}
\epsilon_0 \nabla^2 G_0({\bf x}-{\bf x}') = -\delta({\bf x}-{\bf x'}).
\end{equation}

With this notation the dipole-dipole interaction is given by
\begin{equation}
D_{ij}({\bf x},{\bf x}';L) = \left\{
\begin{array}{cl}
D^{(0)}_{ij}({\bf x}-{\bf x}') & {\bf x}, {\bf x}' \in V^-, V^+\\
D_{ij}^{(0)}({\bf x}-{\bf x}'- L{\bf e}_z) &  {\bf x} \in V^-, {\bf x}' \in V^+\\
D_{ij}^{(0)}({\bf x}-{\bf x}'+ L{\bf e}_z) & {\bf x}' \in V^-, {\bf x} \in V^+.
\end{array}
\right. ~~~
\end{equation}

Note that we have chosen the slabs to be separated by vacuum, so that there is no bulk pressure associated with the intervening dielectric medium, independent of the plate separation $L$. The thermal Casimir force usually given for such systems equals the plate area multiplied by a {\em disjoining pressure}, {\sl i.e.} the difference between the confined and bulk pressures, which tends asymptotically to zero as the plate separation increases.

For a fixed configuration of dipoles (where both their relative position and orientation are fixed) the instantaneous force on $S_2$ is given by
\begin{eqnarray}
\kern-20pt &&f  =- {\partial H \over \partial L}=\nonumber\\ 
&& -\int_{V^-\times V^+}\!\!\!\!\!\!\!\!\!\!\!\!\!\!\!d{\bf x}d{\bf x}' \sum_{ij} p_i({\bf x})\left[ {\partial \over \partial L}D_{ij}^{(0)}({\bf x}-{\bf x}'-L{\bf e}_z)\right]p_j({\bf x}'),~~~~~~~\label{f1}
\end{eqnarray}
as only the interaction between dipoles in different slabs depends on $L$. Of course each dipole feels both the electric fields from dipoles within the same slab and those in the opposing slab. However, when we infinitesimally displace $S_2$, the whole slab moves and each dipole is displaced by the same amount leaving their relative separation and consequently their energy of interaction the same. Let $S_1$ denote the half space $z<0$ and $S_2$ the half space $z>L$, so that the force is given by  
\begin{equation}
\kern-10pt f=- \int_{S_1\times S_2} \!\!\!\!\!\!\!\!\!d{\bf x}d{\bf x}'  \sum_{ij}p_i({\bf x})\left[ {\partial \over \partial z'}D_{ij}^{(0)}({\bf x}-{\bf x}')\right]p_j({\bf x}').\label{f2}
\end{equation}
This expression for the force has an obvious physical interpretation, which could in fact have been used as a starting point for its definition. The electric field due to the dipoles in $S_1$ at the point ${\bf x'}$ in $S_2$ is given by
\begin{equation}
E_{1j}({\bf x'}) = -\int_{S_1} d{\bf x} \sum_{ij}p_i({\bf x})\left[ D_{ij}^{(0)}({\bf x}-{\bf x}')\right],
\end{equation}
and the force in the $z'$ direction on a dipole at ${\bf x}'$ in the $S_2$ due to the dipoles in $S_1$ is thus 
\begin{equation}
F({\bf x'}) =-\sum_j p_j({\bf x'}){\partial\over \partial z'}E_{1j}({\bf x}'),\label{force}
\end{equation}
so Eq. (\ref{f1}) is just $ f = \int_{S_2} d{\bf x'} F({\bf x}')$. Now writing the dipole-dipole interaction in terms of the free Green's function we find  
\begin{equation}
f=- \int_{S_1\times S_2} \!\!\!\!\!\!\!\!\!d{\bf x}d{\bf x}'  \sum_{ij}p_i({\bf x})\left[ {\partial \over \partial z'}\nabla_i\nabla'_jG_0({\bf x}-{\bf x}')\right]p_j({\bf x}').~~~\label{fexp}
\end{equation}
It is important to emphasize here that in this model the electric field is a {\em fixed function} of the dipole 
configurations and that  we do not consider additional thermal fluctuations of the electromagnetic field.

\section{Force fluctuations}
In principle we can compute moments of the force from Eq. (\ref{fexp}), but the calculation can be considerably simplified by considering 
\begin{equation}
{\partial f\over \partial L} =
\int_{S_1\times S_2} \!\!\!\!\!\!\!\!\!d{\bf x}d{\bf x}' \sum_{ij} p_i({\bf x})\left[ {\partial^2 \over \partial z'^2}\nabla_i\nabla'_jG_0({\bf x}-{\bf x}')\right]p_j({\bf x}'),~~~~~~~\label{trick}
\end{equation}
where, as in Eq. (\ref{f2}), we have simply replaced the derivatives in $L$ by derivatives in $z'$. In order to compute the partition function for this system and the correlation function for the dipole field, we first introduce the generating function
\begin{eqnarray}
\kern-10pt &&Z[{\bf u}]=\nonumber \\ 
\kern-10pt &&\int \prod_{i}d[p_i({\bf x})]\exp\left(-\beta H+\!\!\!\int_{S_1\cup S_2} \!\!\!\!\!\!\!\!\!\!\!\!d{\bf x} \ \sum_iu_i({\bf x}) p_i({\bf x})\right),~~~~
\end{eqnarray}
which can be rewritten introducing an auxiliary field $\phi$ that decouples the dipole-dipole interactions. 
We can now integrate over the dipole field to obtain
\begin{eqnarray}
\kern-20pt Z[{\bf u}] &=& \int d[\phi({\bf x})]\exp\left(- {\beta\epsilon_0\over 2}\int d{\bf x} \ [\nabla\phi]^2\right. \nonumber\\
&-&\left.{\beta\over 2}\int_{S_1\cup S_2} \!\!\!\!\!\!\!\!\!d{\bf x}\ \chi({\bf x}) (\nabla\phi({\bf x}) -{i\over \beta}{\bf u}({\bf x}))^2 \right), \label{zu}
\end{eqnarray}
where by setting $\phi=-i\psi$, $\psi$ can be identified as the electrostatic  potential.  Integrating over the dipole field yields the partition function, while the no source term is immediately recognizable as the zero frequency Matsubara mode or thermal (van der Waals) Casimir interaction of the standard Lifshitz theory \cite{par2006}
\begin{equation}
Z[{\bf 0}] = \int d[\phi({\bf x})]\exp\left(- {\beta\over 2}\int d{\bf x} \ \epsilon({\bf x})[\nabla\phi]^2\right)\label{eql}
\end{equation}
with dielectric constants $\epsilon({\bf x}) = \epsilon_0 +\chi_i$ when ${\bf x}$ is in $S_1$ or $S_2$ and 
$\epsilon({\bf x}) =\epsilon_0$ for ${\bf x}$ is between the slabs. Written in the form of Eq. (\ref{eql}), but also from the basic model above, it is clear that our model is valid for high temperature dielectric systems, retardation effects are neglected and also no effects due to conduction electrons is present.

In the absence of sources, the  average value of the dipole field is
$ \langle p_i({\bf x}) \rangle =0$,
and its correlation function at non-coinciding points is
\begin{equation}
\langle p_i({\bf x})p_j({\bf x}') \rangle = -{1\over \beta}\chi({\bf x}) \chi({\bf x}')\nabla_i\nabla_j' G({\bf x},{\bf x}'), \label{corr}
\end{equation} 
where $G$ is the slab geometry Green's function obeying
\begin{equation}
\nabla\cdot\epsilon({\bf x})\nabla G({\bf x},{\bf x}') = -\delta({\bf x}-{\bf x}').
\end{equation} 
The average force is  thermodynamically given  {\sl via}
\begin{equation}
\langle f\rangle = -\langle {\partial H\over \partial L}\rangle = {1\over \beta}{\partial\over \partial L}\ln(Z[{\bf 0}]).
\end{equation}
To compute the variance of the force one notices that 
\begin{equation}
{\partial \over \partial L}\langle f\rangle = -\langle {\partial ^2 H\over\partial  L^2}\rangle + \beta \langle ({\partial H\over \partial L})^2\rangle_c
\end{equation}
which can be rearranged to give the Gibbs lemma \cite{gibbs} type result
\begin{equation}
\langle f^2\rangle_c = k_BT \left[ {\partial \over \partial L}\langle f\rangle
+ \langle {\partial ^2 H\over\partial  L^2}\rangle \right].\label{link}
\end{equation}
The first term on the right-hand-side is easy to calculate. The second term is 
obtained from Eqs. (\ref{trick}) and (\ref{corr}) as 
\begin{eqnarray}
\beta\langle{\partial ^2H\over \partial L^2}\rangle 
= & &-\int_{S_1\times S_2} \!\!\!\!\!\!\!\!\!d{\bf x}d{\bf x}'  \sum_{ij}\left[ {\partial^2 \over \partial z'^2}\nabla_i\nabla'_jG_0({\bf x}-{\bf x}')\right]\nonumber \\ &&\chi({\bf x}) \chi({\bf x}')\nabla_i\nabla_j' G({\bf x},{\bf x}').
\end{eqnarray}
This can be further simplified using the Fourier-Bessel transform of the Green's functions in the ${\bf r}=(x,y)$ plane 
to get  the remarkably simple formula
\begin{equation}
\beta\langle{\partial ^2H\over \partial 
L^2}\rangle= - {A\chi_1\chi_2\over 2\pi}\int dk k^5G(k,0,L)G_0(k,L).
\label{magic}
\end{equation}
The lateral Fourier transform of the free Green's function $G_0$ is given by
\begin{equation}
G_0(k,z-z') = {1\over 2\epsilon_0 k}\exp(-k|z-z'|)
\end{equation}
while
 \begin{equation}
G(k,0,L) = {2\epsilon_0\exp(-kL)\over k(\epsilon_0+\epsilon_1)(\epsilon_0+\epsilon_2)\left(1-\Delta_1\Delta_2\exp(-2kL)\right)},
\end{equation}
where $\Delta_i = (\epsilon_i-\epsilon_0)/(\epsilon_i+\epsilon_0)$.
This finally yields
\begin{equation}
\beta\langle{\partial ^2H\over \partial L^2}\rangle= -{3A\Delta_1\Delta_2\over 8\pi L^4}{\rm Li}_4(\Delta_1\Delta_2),
\end{equation}
where ${\rm Li}_s$ is the polylogarithmic function defined by
\begin{equation}
{\rm Li}_s(z) = \sum_{n=1}^\infty {z^n\over n^s}
\end{equation}
The standard result for the average value of the Casimir force for this system is given by \cite{par2006} 
\begin{equation}
\langle f\rangle = -k_BT A{{\rm Li}_3(\Delta_1\Delta_2)\over 8\pi L^3}.\label{av}
\end{equation} 
Putting these results together in Eq. (\ref{link}) finally yields
\begin{equation}
\langle f^2\rangle_c = {3A(k_BT)^2\over 8\pi L^4}g(\Delta_1\Delta_2)\label{fluc}
\end{equation}
where $ g(z) = {\rm Li}_3(z) -{\rm Li}_4(z)$. The function $g(z)$ is  positive for $z\in[0,1]$, so that indeed the variance of the force fluctuations is everywhere positive. 
We see that the scaling of the fluctuations agrees with the $L$ dependent part in \cite{bar2002}, but contains {\sl no bulk term} dependent on an ultra-violet cut-off. 

The ratio of the variance to the squared average force is then given by
 \begin{equation}
 Q= {\langle f^2\rangle_c\over \langle f\rangle^2}= {24 \pi L^2\over A}h(z)
 \label{fingerprint}
 \end{equation}
 where $h(z)= g(z)/{\rm Li}^2_3(z)$ and $h(z)$ attains its maximum value  $0.08286$ at $z=1$.
 Note the results  are symmetric under the  interchange of the labels $1$ and $2$ and thus the average force on each slab is equal and opposite and the force fluctuations on both slabs are the same. This has to be the case, as from the definition of the unaveraged force it is clear that the forces exerted on each of the slabs are symmetric. 
 \section{conclusions}
We analyzed a simple, classical dipole model, that leads to thermal van der Waals  interactions with the average force  predicted by the standard Lifshitz theory.  In this  classical dipole model   the force at large separations not only goes to zero on the average, but much more stringently with  probability one. The ratio of the variance to the squared average force, Eq. (\ref{fingerprint}), furthermore gives a  {\sl fingerprint} of the classical thermal van de Waals interaction, differentiating it from possible parasitic effects and stray interactions, plaguing experiments in realistic systems \cite{Disorder}.

In our dipole model there is no divergent cut-off dependent bulk contribution to the average force. The fact that no other length scale appears means that the functional form of the force variance in Eq. (\ref{fluc}) can be predicted solely on the grounds of dimensional analysis and by assuming the thermodynamic scaling, {\sl i.e.}  $\Delta f\sim \sqrt{A}$ for the fluctuations as a function of the area of the plates. The so obtained thermodynamic scaling is in agreement with a number of studies of fluctuations of Casimir forces in both thermal and quantum systems \cite{bar1991, ebe1991,bar2002}. The results of \cite{bar2002} are for thermally fluctuating fields, as here, but find a microscopic cut-off  depend force fluctuations at infinite plate separation. Nevertheless the average force has the same behavior in both models. This difference comes naturally from an underlying microscopic definition of the model. In our model the electrostatic fields are generated by the dipoles themselves, being solutions of the Poisson equation for the given dipole distribution. The fields so generated cannot generate self-forces
on isolated objects, hence as the two slabs are separated the total force on each slab decays to zero
and by consequence so does its mean and variance. In the model of \cite{bar2002} (and indeed its quantum counterparts \cite{bar1991, ebe1991}), the fluctuating field exists throughout space  and objects immersed in the field are not its sources. The  objects immersed in the field however modify its fluctuations by the imposition of boundary conditions. Crucially however, the fluctuating field exists in the presence of a single object and hence can exert a force,  albeit of mean zero,  on the object. Intriguingly, the nature of force fluctuations may therefore be relevant to the long standing debate (see \cite{cug12} for a recent discussion) as to the physical interpretation of the Casimir effect in terms of a shift in zero point energy (field effect) or as a van der Waals effect (due to sources). It should be noted that  fluctuations of the force acting on a single particle is compatible with the drag predicted on isolated moving objects coupled to a thermalized fluctuating field \cite{drag,tdrag}.

Normally in electrostatics forces are evaluated by using the stress tensor \cite{lan75} that follows from equations of the form of Eq. (\ref{force}), in order to compute forces generated by electric fields on charges and/or dipoles.  The usual stress tensor given for  dielectric systems depends on the local dielectric constant \cite{lan75} which is already an object derived from  thermodynamic averaging, relating the {\em average} local polarization field to the local  electric field. This means that the validity of the use of the preaveraged stress tensor to compute fluctuations of forces in these types of models is not obvious. The stress tensor before averaging is a fundamental object derived from the force exerted on charges by the electric field. It is indeed this stress tensor that should be used to compute force fluctuations. However, in many coarse grained theories used to describe fluctuating systems, dynamical variables, which are equivalent to charges or dipoles in electrostatic language, have been integrated out in formulating the theory and it is thus not always obvious that they contain sufficient physical information to yield correct predictions for force fluctuations.

\end{document}